\documentclass[proof]{WileyASNA-v1}
\usepackage{url}
\usepackage{makecell}

\articletype{Article Type}%

\received{26 April 2016}
\revised{6 June 2016}
\accepted{6 June 2016}

\begin{document}

\title{Exoplanets in star clusters}

\author[1]{T. Bro{\v z}a*}

\author[1]{E. Paunzen}

\authormark{Bro{\v z}a \& Paunzen}

\address[1]{\orgdiv{Masaryk University}, \orgname{Department of Theoretical Physics and Astrophysics}, \orgaddress{\state{Brno}, \country{Czechia}}}

\corres{*Tobi{\'a}{\v s} Bro{\v z}a, Faculty of Science, Masaryk University, 
Kotl{\'a}{\v r}sk{\'a} 2, 611 37 Brno, Czechia, \email{broza.tobias@gmail.com}}

%\presentaddress{This is sample for present address text this is sample for present address text}

\abstract{The field of exoplanetary research is rapidly evolving, thanks to TESS and Kepler/K2. \textit{Gaia}, with the newest data release \textit{Gaia} DR3, which has transformed the study of star clusters. By combining the known properties of exoplanets with those of the star clusters in which they are found, it is possible to study the planets in greater detail and test theories of planetary evolution.
We aim to create a catalogue of \textit{Gaia} IDs for known exoplanets and use it to link the planets to a catalogue of star clusters. We utilised the NASA Exoplanet Archive and the Extrasolar Planets Encyclopaedia to compile a catalogue of known exoplanets. We then cross-matched the data with sources in \textit{Gaia} DR3 to obtain the \textit{Gaia} IDs of the exoplanets. Using this information, we inferred the potential association of these exoplanets with star clusters. 
We described the statistical properties of these exoplanets. We compiled a catalogue of \textit{Gaia} IDs for known exoplanets and identified those associated with star clusters. This resource facilitates detailed studies of exoplanet properties and their evolutionary contexts within stellar clusters.}


%\jnlcitation{\cname{%
%\author{Williams K.}, 
%\author{B. Hoskins}, 
%\author{R. Lee}, 
%\author{G. Masato}, and 
%\author{T. Woollings}} (\cyear{2016}), 
%\ctitle{A regime analysis of Atlantic winter jet variability applied to evaluate HadGEM3-GC2}, %\cjournal{Q.J.R. Meteorol. Soc.}, \cvol{2017;00:1--6}.}

\fundingInfo{Funding info text.}

\maketitle

%\footnotetext{\textbf{Abbreviations:} ANA, anti-nuclear antibodies; APC, antigen-presenting cells; IRF, interferon regulatory factor}

\section{Introduction}\label{}

An exoplanet or extrasolar planet is a planet outside the solar system. The first possible evidence of an exoplanet 
was noted by \citet{1917PASP...29..258V} but was not then recognized as such. First confirmed detection came in 1992 \citep{wolszczan_planetary_1992}, and as of October 2024, the existence of more than 5800 exoplanets has been confirmed \citep{ps}. Planets vary greatly in size, ranging from rocky planets smaller than Mercury \citep[$\approx0.00017\text{ M}_\text{Jup}$,][]{barclay_sub-mercury-sized_2013} to giant planets of theoretical mass up to around $60\text{ M}_\text{Jup}$ \citep{hatzes_definition_2015}. Discoveries of exoplanets, mostly by missions TESS \citep{ricker_transiting_2014} and Kepler/K2 \citep{borucki_kepler_2010}, accompanied by other ground-based observations, led to a significant shift in planetary theories, which were previously centred solely around the history and evolution of the solar system.\par

Star clusters are a group of gravitationally bound stars which formed from the same molecular cloud, meaning they share certain characteristics, e.g., age, metallicity, distance, etc., thus providing an ideal environment for the study of star formation and evolution \citep{krumholz_star_2019}. Thanks to their properties, we can use star clusters in order to study exoplanets in more depth, since it is possible to assume that the metallicity and age of a planet correspond to those of a star cluster \citep{teske_starplanet_2024}. Thus, studying clusters enables us to gain a deeper understanding of exoplanets. Furthermore, data from Kepler show that around 30 $\%$ of stars host at least one planet \citep{zhu_exoplanet_2021}. In addition, \citet{lada_embedded_2003} claim that the percentage of stars that form in clusters is likely in the range of 70-90 $\%$.\par

\textit{Gaia} is a satellite of the European Space Agency, which was launched in 2013 and orbits the Sun-Earth L$_2$. The satellite carries instruments, which enable us to gather unprecedentedly precise astrometric and photometric measurements \citep{brown_gaia_2016}. Compared to its predecessor, the \textit{Hipparcos}, \textit{Gaia} offers an order of magnitude improvement in parallax and proper motion measurements for $10^4$ times as many sources as \textit{Hipparcos}, totalling more than 1.8 billion. \textit{Gaia} works by scanning the entire sky, providing us with a 3D map. The satellite has proven influential, as showcased by works such as \citet{bossini_age_2019}, who utilised \textit{Gaia} DR2 to determine ages, distances and extinctions for 269 open clusters (OCs). Similarly, \citet{meingast_extended_2021} have used \textit{Gaia} DR2 to study chosen nearby young clusters further, while \citet{tarricq_structural_2022} applied a clustering algorithm to \textit{Gaia} EDR3, identifying 389 OCs and determining their structural parameters.\par 

This data was then used by \citet{hunt_improving_2023} to create an unbiased and precise catalogue of globulars and OCs. They achieved this by using $\textit{Gaia}$ DR3\footnote{\textit{Gaia} DR3 is the latest data release. It was released in June 2023.}. They used the Hierarchical Density-Based Spatial Clustering of Applications with Noise (HDBSCAN) algorithm \citep{campello_density-based_2013}. However, a major flaw of HDBSCAN is its high false positive rate \citep{hunt_improving_2021}. Hunt \& Reffert overcome this obstacle by training a neural network to validate members' photometric data. Overall, they recovered 7,167 clusters, 2,387 of which were new candidate objects, with the remainder cross-matched to objects in the literature.\par

The first exoplanet orbiting a cluster member was discovered by \citet{sato_planetary_2007}. The star, $\epsilon$ Tau, is located in the Hyades star cluster. TESS and Kepler/K2 mentioned above were of great importance in this regard. Several projects focused on exoplanets in star clusters. The Zodiacal Exoplanets in Time (ZEIT) project utilises K2 data to search for planets in nearby young clusters, allowing for the study of planets in their most formative years. ZEIT has found planets in Hyades, Pleiades, Upper Scorpius, Praesepe \citep{mann_zodiacal_2016a,mann_zodiacal_2016b,mann_zodiacal_2017a,mann_zodiacal_2017b,gaidos_zodiacal_2017,gaidos_zodiacal_2021,rizzuto_zodiacal_2018,vanderburg_zodiacal_2018}. \citet{bouma_cluster_2019} describes a Cluster Difference Imaging Photometric Survey, which aims to discover giant transiting planets with known ages and to extract light-curves of stars. This resulted in the finding of an exoplanet in the open cluster IC 2602 \citep{bouma_cluster_2020}. Furthermore, \citet{bouma_38_2022,bouma_kepler_2022} describe the discovery of exoplanets in the Cep-Her complex, more specifically in subclusters $\delta$ Lyr, CH-2 and RSG-5.  Likewise, The GAPS programme at TNG has discovered several planets in M44 \citep{malavolta_gaps_2016}. Another project that has detected exoplanets in star clusters, the TESS Hunt for Young and Maturing Exoplanets collaboration (THYME) \citep{newton_tess_2019}, has done so in the Tucana-Horologium association \citep{newton_tess_2019}, the Sco-Cen association \citep{rizzuto_tess_2020}, the Ursa Major group \citep{mann_tess_2020,capistrant_tess_2024}, the Lower Centaurus Crux \citep{mann_tess_2022}, the young cluster Group-X \citep{newton_tess_2022}, the Hyades cluster \citep{distler_tess_2024}, and in association the collaboration has found, Melange-1, Melange-3 and Melange-5 \citep{tofflemire_tess_2021,barber_transit_2022,thao_tess_2024}. Building on this momentum, the TESS Investigation—Demographics of Young Exoplanets (TI-DYE) survey has focused on identifying planets in the youngest stellar associations ($<$50 Myr). Notable discoveries from this survey include one of the youngest known transiting planets, IRAS 04125+2902 b \citep{barber_giant_2024}, and the validation of several Jupiter-sized planets in extremely young groups like HIP 67522 and TOI-6448 and detecting TOI-2076 e \citep{barber_tess_2024, barber_tess_2025,barber_tess_2025-1}. Moreover, \citet{nardiello_psf-based_2019,nardiello_psf-based_2020,nardiello_psf-based_2021,nardiello_psf-based_2020-1} of the project PSF-based Approach to TESS High-quality Data of Star Clusters (PATHOS) used data from the TESS mission and managed to recover light-curves of 90 planetary candidates. Similarly, \citet{fernandes_pterodactyls_2022} introduced the \texttt{pterodactyls} pipeline, specifically designed to search TESS data for transiting exoplanets, and it has been successfully tested on several clusters. They have also looked into the potential shrinkage of the planets due to atmospheric loss throughout time. In addition, \citet{dai_understanding_2023} searched for exoplanets and planetary candidates in OCs and moving-group catalogues and claims to have found 73 confirmed planets and 84 candidates. They use $Gaia$ DR2 and exoplanet catalogues from May 2023, meaning their data can no longer be considered outdated, thanks to $Gaia$ DR3. Recent statistical analyses, such as those by \citet{vach_occurrence_2024}, have utilised TESS-discovered young samples to show that planets younger than 200 Myr are significantly larger than the older population discovered by Kepler, likely due to thermal contraction or atmospheric loss. This highlights the necessity of precise cluster-based catalogues to track the radius evolution of planets over cosmic time. Another project focusing on exoplanets in star clusters is The Young Exoplanet Transit Initiative (YETI) \citep{neuhauser_young_2011}. YETI has observed stars in Trumpler 37, 25 Ori, NGC 7243 and IC 348. \par

Here we describe the creation of a catalogue of \textit{Gaia} IDs of discovered exoplanets and candidate objects. This catalogue can be used in the future to further study planets using astrometric, photometric and spectroscopic data measured by \textit{Gaia}. In this work, we use this list to link it to a catalogue of star clusters, moving groups, and stellar associations compiled by \citet{hunt_improving_2023}. We describe the creation of the catalogue and the properties of exoplanets in the star clusters.

Future satellite missions like PLATO \citep{2014ExA....38..249R} and the Nancy Grace Roman 
Space Telescope \citep{2021A&A...651A...7C} will also hunt for exoplanets in star clusters. Other than the 
TESS mission \citep{2021ApJS..254...39G}, the pixel resolution will allow for unambiguously identifying 
transits and their host stars in crowded fields. Our catalogue will help to lay the groundwork for preparing
observations in star clusters.

\begin{figure}[t]
    \centering
    \includegraphics[width=0.9\linewidth]{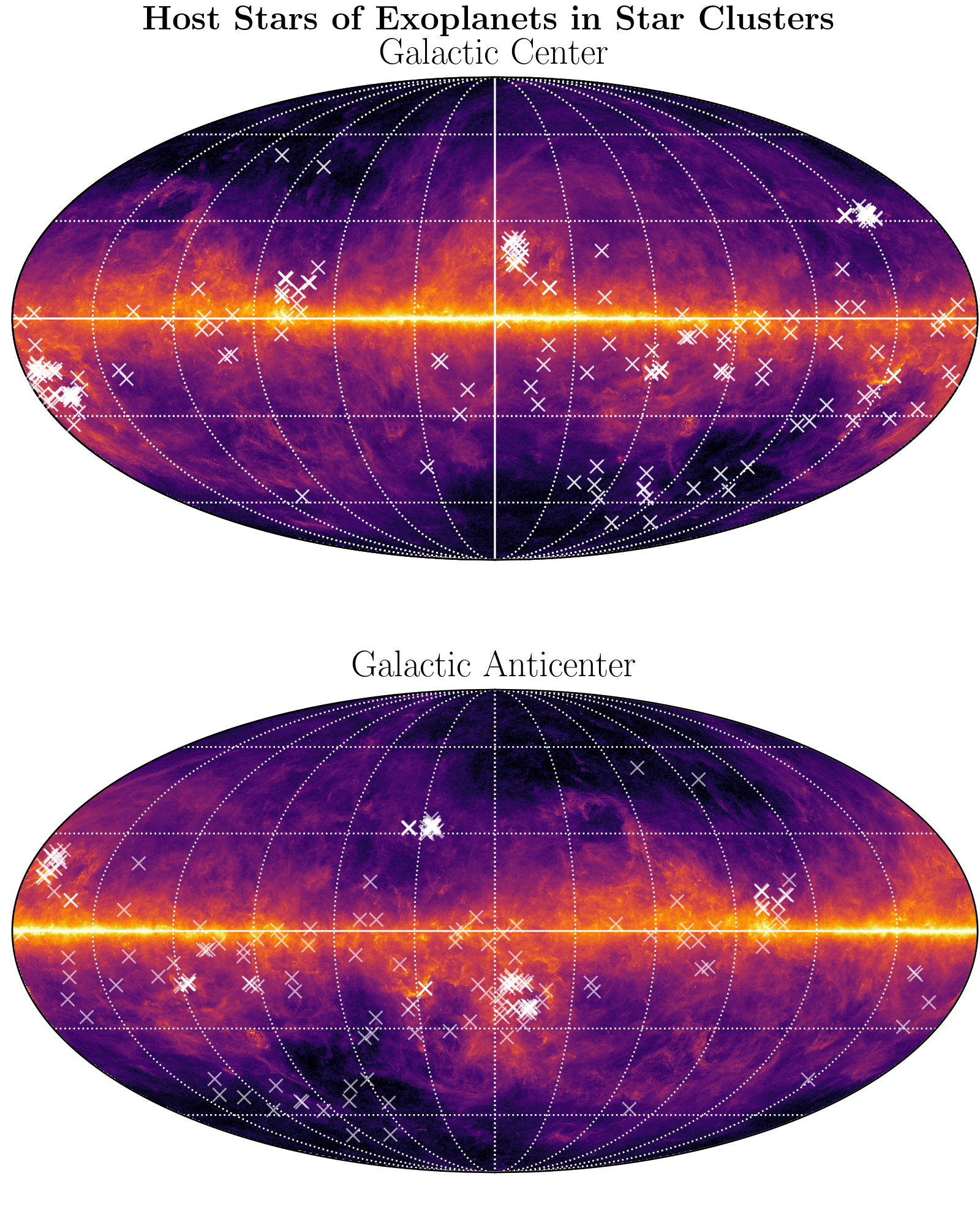}
    \caption{Sky maps with locations of star clusters with exoplanets and candidate objects. The upper graph is centred on the Galactic centre, while the lower one is centred on the Galactic Anticenter.}
    \label{fig:positions}
\end{figure}

\begin{figure}[t]
    \centering
    \includegraphics[width=0.9\linewidth]{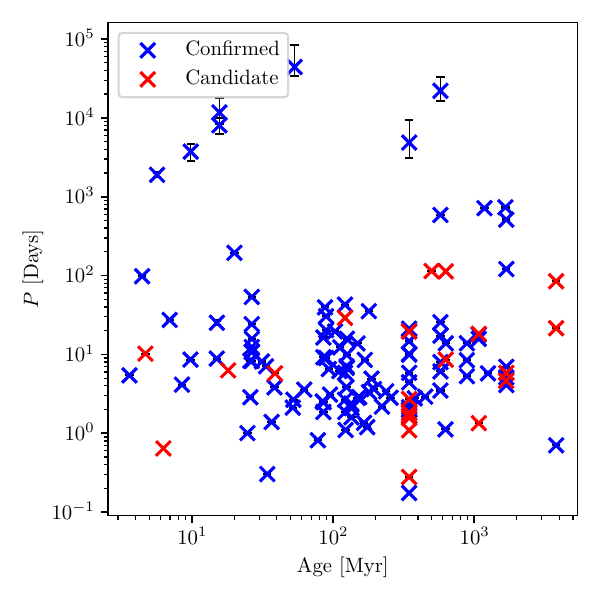}
    \caption{A log-log scatter plot of orbital period $P$ of an exoplanet plotted against the age of its host cluster in Myr, with differentiated confirmed exoplanets (blue) and candidate objects (red). It is possible to argue that older planets tend to have shorter orbital periods; however, a larger sample size would be needed.}
    \label{fig:period_age}
\end{figure}

\section{Method} \label{method}

To describe exoplanets in a star cluster, it is crucial to figure out which planets belong to which clusters. We achieved this by creating a catalogue of known exoplanets and the $Gaia$ ID \footnote{Unique identifier assigned to every object observed by \textit{Gaia}} of their host star. To create the catalogue, we utilised the coordinates of known exoplanets. These data were taken from NASA Exoplanet Archive and The Encyclopaedia of Exoplanetary Systems\footnote{Available at \url{https://exoplanetarchive.ipac.caltech.edu/index.html} and \url{https://exoplanet.eu/home/, respectively}.}.

\subsection{NASA Exoplanet Archive}

The NASA Exoplanet Archive offers an up-to-date list of known exoplanets and their properties. Its infrastructure was adapted from the NASA Stellar and Exoplanet Database. It includes exoplanets, whose discovery has been made available in a peer-reviewed publication, the exoplanets minimum mass is equal to or less than 30 $M_{Jup}$. It also does not include free-floating planets and sufficient follow-up observations, and validation is required, in order to minimise the possibility of a false positive \citep{ExoFOP2019-op,NASA_Exoplanet_Archive2019-xi,NASA_Exoplanet_Science_Institute2020-ir,Ciardi2021-us}. In this work, we used data from the 18th of March 2025, when it comprised 5856 confirmed planets. The Exoplanet Archive also includes candidate objects from the TESS, Kepler and K2 missions. These were included in order to maximise the number of entries.

\subsection{The Encyclopaedia of Exoplanetary Systems}

The Extrasolar Planets Encyclopaedia is an evolving online catalogue, meaning it is updated as new exoplanets are discovered. It was established in 1995. For a planet to be included in the catalogue, it needs to fulfil several criteria. The basic criterion is that the mass does not exceed 60 $M_{Jup}+1\sigma$. Furthermore, the planet discovery must be published in a professional journal, submitted to a professional conference, or announced at a professional conference. Data on the planets are taken from the latest published papers, professional preprints, conference proceedings, and professional websites. The catalogue also includes information about the planets host star \citep{martin_encyclopaedia_1995}. The data from the catalogue were taken on the 18th of March 2025, when it contained information about 7\,427 exoplanets. 

\subsection{The catalogue} \label{catalogue}

Using the equatorial coordinates of the exoplanets from the catalogues described in Sect. \ref{method}, we searched for objects in $Gaia$ DR3 in the online database Vizier\footnote{https://vizier.cds.unistra.fr/viz-bin/VizieR-3?-source=I/355/gaiadr3}. A search radius of 5 arcsec was chosen, after which another search was conducted with coordinates for which no objects were found. This time with a search radius of 10 arcsec to include objects with high proper motion. \par

Naturally, alongside the source identifiers of exoplanet host stars, the search returned identifiers of other objects in a five-arcsec (10-arcsec) radius. Using other parameters measured by \textit{Gaia}, mostly the G-band mean magnitude, which was compared with known magnitudes of host stars, it was determined which object in the DR3 is the actual host star. In cases where it was not possible to decide using magnitude, an astronomical database SIMBAD\footnote{https://simbad.unistra.fr/simbad/} was used. \par

Host stars for which the correct source identifier could not be identified were not included in the final database. The database contains more than 14 000 unique \textit{Gaia} source identifiers of host stars of exoplanets and candidate objects.

\subsection{Exoplanets in star clusters} \label{exoplanets in clusters}

Using the \textit{Gaia} IDs of host stars (Sect. \ref{catalogue}), it is possible to match them with stars belonging to clusters found by \citet{hunt_improving_2023}. This creates a list of star clusters that contain an exoplanet. Since molecular clouds collapse, which cause star formation, happen relatively quickly (in the star formation scale) and thus all stars in a star cluster were created at roughly the exact moment ($\pm$ the timescale of star formation), we can assume with relatively high accuracy that the age of a cluster corresponds to the age of a planet \citep{liu_understanding_2024,parker_birth_2020,richert_circumstellar_2018}. 

\begin{figure}[t]
    \centering
    \includegraphics[width=0.9\linewidth]{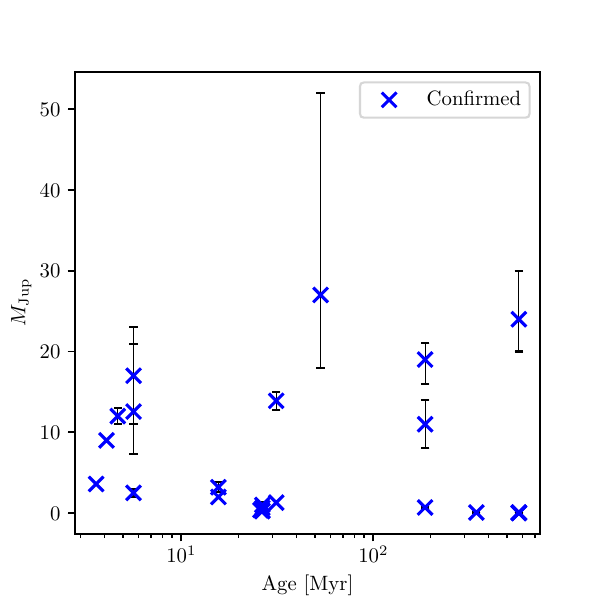}
    \caption{A scatter plot of exoplanets' masses (in units of Jupiter masses) plotted against the age of their host cluster in Myr.}
    \label{fig:mass_age}
\end{figure}

\section{Results} \label{results}

\begin{figure}[t]
    \centering
    \includegraphics[width=0.9\linewidth]{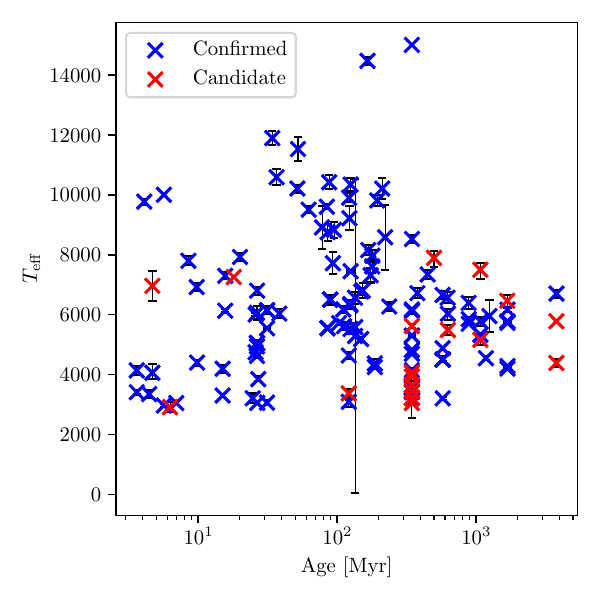}
    \caption{A scatter plot of effective stellar temperatures of host stars of exoplanets or candidate objects plotted against the age of their host cluster in Myr.}
    \label{fig:stellar_temp_age}
\end{figure}

\begin{figure}[t]
    \centering
    \includegraphics[width=0.9\linewidth]{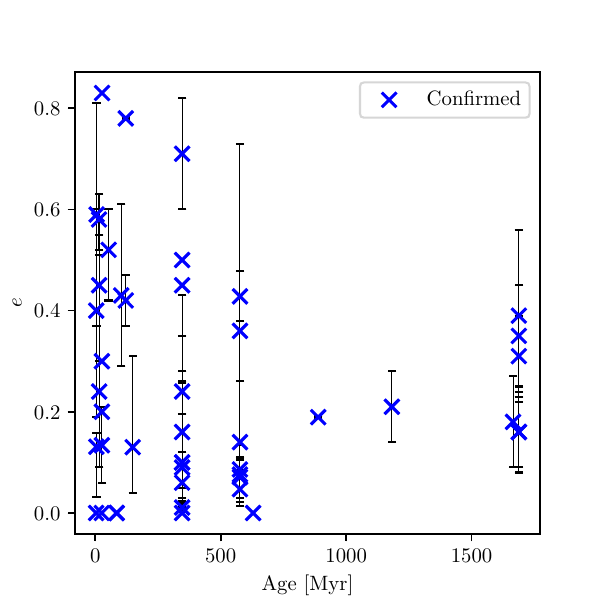}
    \caption{A scatter plot of eccentricities of exoplanetary orbits plotted against the age of their host clusters in Myr.}
    \label{fig:eccentricity_age}
\end{figure}

For the statistical analysis, we used Pearson correlation coefficients and Student's t-test
\citep{1996asst.book.....B}. It has to be emphasised that we are in a poor number regime, which makes
any statistical analysis difficult.
Overall, we found 229 host stars of exoplanets or candidate objects, whose positions are shown in Figure \ref{fig:positions}, inside 89 clusters (Table
\ref{table:list_stars}). An interesting finding is the age range of the clusters, with the youngest being 3.6 Myr and the oldest 3\,800 Myr. Furthermore, it is possible to argue that older planets tend to have shorter orbital periods, but a larger sample size would be needed (Figure \ref{fig:period_age}). In addition, we used accurately determined masses of the planets. Unfortunately, the sample size is even smaller, but it is possible to argue that younger planets tend to have lower mass (Figure \ref{fig:mass_age}). Additionally, the relationship between the effective stellar temperature of exoplanet host stars and their age is plotted in Figure \ref{fig:stellar_temp_age}. The dependence of eccentricity on its host stars' age is visible in \ref{fig:eccentricity_age}. \par

Moreover, our data suggest specific clusters that appear highly relevant to future research. Table \ref{table:ages} showcases some of the youngest and oldest clusters, which contain an exoplanet or a candidate object. One of the notable clusters is M44, where we identified 12 stars that host an exoplanet. This represents the highest number of exoplanets in a single cluster in our data. Another noteworthy cluster, M67, with five host stars, has been extensively studied.  Furthermore, the Orion Complex is currently being studied by many researchers. Similarly to Chameleon I, the Ophiuchus complex contains many clusters, where stars are being born. For this reason, we find the youngest exoplanets in these two.

To estimate how many more data points are needed to significantly improve the statistics, 
we applied a detailed Monte Carlo simulation \citep{2012A&A...537A..67B} using our data set. 
We generated synthetic data sets using our calculated significance levels and extended them up
to ten times the sample size given here. It shows that an increase of a factor of 1.5 would 
allow us to make definite conclusions about the discussed correlations. Let's keep in mind that 
TESS severely suffers from the large pixel size of its detector. Therefore, we can only expect
to find additional transits in very close open clusters. However, the first long-pointing field
of the PLATO mission will include a significant number of open clusters \citep{2025A&A...694A.313N}.

\begin{table}
\caption{The list of matched stars hosting an exoplanet.}                 % title of Table
\label{table:ages}    % is used to refer this table in the text
\centering                        % used for centering table
\begin{tabular}{c c c}      % centered columns (3 columns)
\hline\hline               % inserts double horizontal lines
Cluster & $n$ & Age [Myr] \\         % table heading
\hline                      % inserts single horizontal line
   OCSN\_100         & 4 & 3.6 \\    % inserting body of the table
   $\sigma$ Orionis & 4 & 3.7 \\
   Theia\_54         & 1 & 4.1 \\
   OCSN\_98          & 1 & 4.5 \\
   OCSN\_96          & 7 & 4.7 \\
   HSC\_1318         & 4 & 5.5 \\
   Chamaleon\_I      & 11 & 5.7 \\
\hline
   NGC\_6791         & 2 & 3800 \\
   Theia\_6046       & 1 & 3800 \\
\hline                                  %inserts single line
\end{tabular}
\end{table}

\section{Discussion} \label{discussion}
A catalogue containing \textit{Gaia} identifiers can significantly aid in the study of exoplanets, as it enables researchers to access information about the planet's host star, which can then be used to study the planet further, since the coordinates in the catalogue can be used to trace the object back to exoplanet catalogues containing additional parameters of the planetary system. Such a list has never been created before; thus, a large-scale study of exoplanets using \textit{Gaia} data is the first of its kind. However, this catalogue has several drawbacks. Firstly, the primary identifying criterion was a star's magnitude; however, the $G$-magnitude can differ significantly from magnitudes measured by other instruments. This means the catalogue is likely to contain errors. The list is also incomplete because, in some instances, we were unable to confidently assign the correct \textit{Gaia} ID. In addition, automating this process would be beneficial, as it could react to new data, unlike this catalogue, which was created manually. Another important consideration is the difference between the reported ages. In this work, we use the ages from \citet{hunt_improving_2023}, which are derived using a homogeneous methodology based on fitting isochrones, whereas the ages reported in exoplanet catalogues may be determined using a variety of methods and assumptions. \par

\begin{figure}
    \centering
    \includegraphics[width=0.9\linewidth]{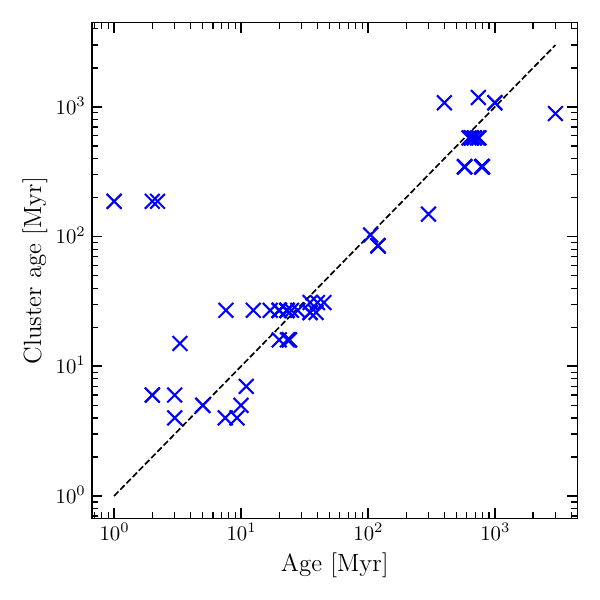}
    \caption{Comparison of cluster ages reported by \citet{hunt_improving_2023} and ages from the NASA Exoplanet Archive. The dashed line denotes the 1:1 relation.}
    \label{fig:age}
\end{figure}

Future works would benefit from more data. The next \textit{Gaia} Data Release (DR4) is scheduled for release in late 2026, and DR5 is expected in 2030 \citep{noauthor_gaia_nodate}. This will significantly increase the accuracy of the data, leading to new research, a deeper understanding of star clusters, and the discovery of new ones that could contain exoplanets. Moreover, several exoplanets were discovered using the $Gaia$ data \citep{arenou_gaia_2023}. Naturally, the discovery of more exoplanets and the accurate determination of their properties would improve this work. There are several projects aimed at discovering and studying exoplanets, the most notable of which is the Planetary Transits and Oscillations of Stars (PLATO) mission, scheduled to launch in 2026. PLATO will be equipped with 26 cameras to monitor stars for extended periods. PLATO aims to discover exoplanets using transits, transit timing variations, and reflected light. PLATO is designed to provide insight into planets' masses, which is relevant to future work \citep{PLATOmission_esa_2018}. The Nancy Grace Roman telescope is set to launch in 2027. Alongside its work on dark matter and contributions to infrared astrophysics, this telescope aims to aid the study of exoplanets. It will be able to detect planets via microlensing events, making it feasible to measure their masses and orbital elements \citep{miyazaki_revealing_2021}. Detections using transits \citep{tamburo_predicting_2023} and direct imaging \citep{mennesson_roman_2021} should also be possible. Combined with other ground-based observations, such as those from the Extremely Large Telescope, which is currently under construction and will be able to discover exoplanets, determine their masses, radii, and compositions, study circumstellar and planetary disks, and observe star-forming regions \citep{elt_programme_2011}, this will significantly improve the exoplanet census. Furthermore, it is worth noting that the list of exoplanets is likely significantly biased, primarily towards heavier planets with shorter orbits, which is attributed to the effectiveness of the transit technique \citep{kipping_observational_2016}.

\section{Conclusions} \label{conclusions}
In this work, a catalogue containing the $Gaia$ identifiers of parent stars of known exoplanets and planet candidates was created, and will be made available. Thanks to the accuracy and amount of data from the $Gaia$ mission, the identifiers are helpful for further analysis and study of exoplanets. \par

Using the catalogue, these parent stars have been matched to star clusters found by \citet{hunt_improving_2023}. Thanks to the properties of star clusters, it is possible to determine their age accurately. This is the first work to focus on exoplanets in star clusters of this scale. Using cluster ages, an analysis of planetary properties was conducted. In this work, we focused on the spatial distribution of exoplanets, as well as their orbital periods, eccentricities, and masses. Star clusters containing an exoplanet seem to be evenly spread out in the Milky Way and reflect the positions of known clusters. Younger planets may have longer orbital periods, but more thorough statistical analysis is needed to confirm this. Eccentricity does not seem to be influenced by the age of its parent star, which is relevant for modelling solar systems. Younger planets may be lighter, but for most planets, only the lower limit of their mass is known, so the pool of planets with accurately determined mass is insufficient. The data also show that planets found in star clusters are relatively young, and candidate planets from the TESS mission generally orbit stars with higher effective temperatures. \par

Furthermore, we were interested in whether the effective stellar temperature plays a role in planetary formation. There seem to be envelopes constraining the effective stellar temperature based on the age of the star. While there does not seem to be a relation between a planet's eccentricity and the effective temperature of its host star, there is a possible correlation between a planet's orbit and the stellar temperature of the star around which it orbits. \par
The catalogue can be helpful for further study of exoplanets. The data hint at some possible correlations, but a larger, less biased sample of exoplanets, combined with careful statistical analysis, is needed. 
This work can serve as a basis for future studies of exoplanets and their host stars, using the star cluster to which they belong.

\section*{Acknowledgments}

This work was supported by the grant \fundingAgency{GAČR} \fundingNumber{23-07605S} and was carried out within the institutional support framework 
for the development of the
research organization of Masaryk University.
      Based partly on data acquired at the Anglo-Australian Telescope in the semester 2009A. We acknowledge the traditional custodians of the land on which the AAT stands, the Gamilaraay people, and pay our respects to elders past and present.
      This work presents results from the European Space Agency (ESA) space mission 
      Gaia. Gaia data are being processed by the Gaia Data Processing and Analysis 
      Consortium (DPAC). Funding for the DPAC is provided by national institutions, 
      in particular, the institutions participating in the Gaia Multilateral Agreement 
      (MLA). The Gaia mission website is https://www.cosmos.esa.int/gaia. 
      The Gaia archive website is https://archives.esac.esa.int/gaia.
      This publication makes use of data products from the Two Micron All Sky Survey, which is a joint project of the University of Massachusetts and the Infrared Processing and Analysis Center/California Institute of Technology, funded by the National Aeronautics and Space Administration and the National Science Foundation.
      This research has made use of the WEBDA database, operated at the Department of Theoretical Physics and Astrophysics of Masaryk University.

%\subsection*{Author contributions}

%This is an author contribution text. This is an author contribution text. This is an author %contribution text.  

%\subsection*{Financial disclosure}

%None reported.

\subsection*{Conflict of interest}

The authors declare no potential conflict of interest.

\subsection*{Data Availability Statement}

The data that support the findings of this study are available from the corresponding author upon reasonable request.

\appendix

\section{The list of matched stars hosting an exoplanet\label{app1}}

\begin{table*}[t]
\caption{The list of matched stars hosting an exoplanet together with the corresponding cluster and their age in Myr.}                 % title of Table
\label{table:list_stars}    % is used to refer this table in the text
\centering                        % used for centering table
\begin{tabular}{c c c c c c c}      % centered columns (3 columns)

\hline               % inserts double horizontal lines
Gaia ID Star & Right Ascension & Declination & Cluster & Age [Myr] & Discovery Method & Status\\         % table heading
\hline                      % inserts single horizontal line
2741090498161113344	&	3.987500	&	+04.251111	&	HSC\_749	&	226 & Imaging & Confirmed	\\
4901229043960053248	&	4.608333	&	$-$63.477500	&	beta\_Tuc\_Group	&	31 & Astrometry & Confirmed	\\
427614644875238528	&	10.962717	&	+61.835626	&	NGC\_225	&	93	& Transit & Confirmed\\
4903834989597325312	&	12.920080	&	$-$59.225578	&	beta\_Tuc\_Group	&	31 &  Imaging & Confirmed	\\
4954323704550180352	&	25.492637	&	$-$46.565930	&	beta\_Tuc\_Group	&	31 & Imaging & Confirmed\\
4963614887043956096	&	34.842705	&	$-$39.423077	&	beta\_Tuc\_Group	&	31 & Imaging  & Confirmed	\\
4738287468036620160	&	35.977725	&	$-$58.251842	&	beta\_Tuc\_Group	&	31 & Imaging, Astrometry  & Confirmed	\\
4738227407213866368	&	36.331155	&	$-$58.624891	&	beta\_Tuc\_Group	&	31 & Imaging, Astrometry & Confirmed	\\
4744825404693448320	&	36.735824	&	$-$53.450879	&	beta\_Tuc\_Group	&	31 & Astrometry  & Confirmed	\\
338236787755719936	&	37.132410	&	+40.891254	&	UPK\_303	&	111 & Transit &	Confirmed\\
4697085327076101760	&	38.503899	&	$-$64.701960	&	beta\_Tuc\_Group	&	31 & Imaging & Confirmed	\\
466007701130563200	&	39.891102	&	+62.433306	&	HSC\_1103	&	63 & Transit  & Confirmed	\\
337169505562029568	&	40.245588	&	+42.871248	&	NGC\_1039	&	122 & Transit  & Confirmed	\\
4846216555917018624	&	50.791710	&	$-$46.523217	&	beta\_Tuc\_Group	&	31 & Imaging & Confirmed	\\
4860787013426321664	&	52.647353	&	$-$35.203715	&	Alessi\_13	&	25 &  Transit & Confirmed	\\
119520079331712896	&	54.183897	&	+28.550269	&	Melotte\_22	&	122 & Transit  & Confirmed	\\
56950239849717248	&	54.397778	&	+18.896264	&	Melotte\_22	&	122 & Transit  & Candidate 	\\
65048589663795200	&	55.475654	&	+23.084738	&	Melotte\_22	&	122 & Imaging & Confirmed	\\
37619725922094336	&	55.633333	&	+12.272778	&	Melotte\_25	&	577 & Transit  & Confirmed	\\
69935678330276992	&	56.096809	&	+25.645871	&	Melotte\_22	&	122 & Imaging & Confirmed	\\
216678286879869056	&	56.141834	&	+32.115858	&	IC\_348	&	6 & Imaging & Confirmed	\\
69816102146573440	&	56.380662	&	+24.879903	&	Melotte\_22	&	122 & Imaging  & Confirmed	\\
65229008357099136	&	56.461075	&	+24.151030	&	Melotte\_22	&	122 & Imaging  & Confirmed	\\
65224889481112832	&	56.608695	&	+24.086012	&	Melotte\_22	&	122 &  Imaging & Confirmed	\\
64808204638390912	&	56.659968	&	+22.919778	&	Melotte\_22	&	122 &  Imaging & Confirmed	\\
66728265476081408	&	56.730555	&	+24.187804	&	Melotte\_22	&	122 & Transit  & Confirmed	\\
66734720809017856	&	56.825000	&	+24.390833	&	Melotte\_22	&	122 & Imaging & Confirmed	\\
64005183194070272	&	56.865505	&	+22.160810	&	Melotte\_22	&	122 & Imaging & Confirmed	\\
66791521754962560	&	56.912541	&	+24.606221	&	Melotte\_22	&	122 &  Imaging & Confirmed	\\
65000073712701056	&	57.020833	&	+23.658333	&	Melotte\_22	&	122	& Radial Velocity & Confirmed\\
66738328582414720	&	57.079167	&	+24.420278	&	Melotte\_22	&	122 & Imaging  & Confirmed	\\
66765855029900288	&	57.186202	&	+24.623230	&	Melotte\_22	&	122 &  Imaging & Confirmed	\\
64127881818726400	&	57.210139	&	+22.741681	&	Melotte\_22	&	122 & Imaging  & Confirmed	\\
66498158310444288	&	58.027952	&	+24.266845	&	Melotte\_22	&	122 & Imaging  & Confirmed	\\
65644451951851776	&	58.479647	&	+23.393417	&	Melotte\_22	&	122 & Imaging & Confirmed	\\
4841448081361281920	&	59.362339	&	$-$44.291815	&	beta\_Tuc\_Group	&	31 & Imaging & Confirmed	\\
51369428065705600	&	59.791667	&	+20.160000	&	HSC\_1340	&	27 & Imaging  & Confirmed	\\
51886335968692480	&	61.331654	&	+20.157032	&	HSC\_1340	&	27 & Transit  & Confirmed	\\
51914888911355776	&	61.662500	&	+20.303056	&	HSC\_1340	&	27 & Imaging & Confirmed	\\
45159901786885632	&	61.755663	&	+15.334944	&	Melotte\_25	&	577 & Radial Velocity  & Confirmed	\\
4844691297067063424	&	62.966385	&	$-$37.939730	&	HSC\_1923	&	85 & Transit  & Confirmed	\\
3311804515502788352	&	63.273935	&	+15.247703	&	Melotte\_25	&	577 & Transit  & Confirmed	\\
163182888662060928	&	63.549517	&	+28.198169	&	HSC\_1318	&	6 & Imaging & Confirmed	\\
163165738856771200	&	63.811340	&	+28.002607	&	HSC\_1318	&	6 & Imaging, Other  & Confirmed 	\\
164800235906366976	&	63.928339	&	+29.166542	&	CWNU\_1129	&	15 &  Transit & Confirmed	\\
164495323291866624	&	64.713135	&	+28.242583	&	HSC\_1318	&	6 &  Imaging, Other & Confirmed	\\
47316005432912256	&	64.715445	&	+17.387883	&	HSC\_1340	&	27 &  Radial Velocity & Confirmed	\\
152516079683687680	&	65.775330	&	+28.022097	&	HSC\_1318	&	6 & Imaging & Confirmed	\\
\hline                                  %inserts single line
\end{tabular}
\end{table*}

\addtocounter{table}{-1}

\begin{table*}[t]
\caption{continued.}                 % title of Table
\label{table:list_stars}    % is used to refer this table in the text
\centering                        % used for centering table
\begin{tabular}{c c c c c c c}      % centered columns (3 columns)

\hline               % inserts double horizontal lines
Gaia ID Star & Right Ascension & Declination & Cluster & Age [Myr] & Discovery Method & Status \\         % table heading
\hline                      % inserts single horizontal line
149629483705467008	&	65.897496	&	+25.050671	&	Theia\_7	&	187 & Imaging  & Confirmed	\\
48026706558487040	&	67.154657	&	+19.180278	&	Melotte\_25	&	577 &  Radial Velocity & Confirmed	\\
145916050683920128	&	67.412862	&	+22.882567	&	Melotte\_25	&	577 & Transit  & Confirmed	\\
151374202498079872	&	67.423167	&	+26.549390	&	Theia\_7	&	187 &  Imaging & Confirmed	\\
151102790629500288	&	67.738349	&	+25.944319	&	Theia\_7	&	187 & Imaging, Other & Confirmed	\\
3285426613077584384	&	68.020521	&	+05.410075	&	Melotte\_25	&	577 & Imaging  & Confirmed	\\
147799209159857280	&	68.074465	&	+24.370806	&	Theia\_7	&	187 & Imaging  & Confirmed	\\
3314132593936245248	&	68.126153	&	+17.525125	&	Theia\_66	&	6 & Imaging  & Confirmed	\\
147801339463632000	&	68.258215	&	+24.350020	&	Theia\_7	&	187	& Other & Confirmed\\
147831571737487488	&	68.291826	&	+24.561924	&	Theia\_7	&	187 & Radial Velocity  & Confirmed	\\
145203159127518336	&	68.466769	&	+22.841619	&	CWNU\_1129	&	15 & Radial Velocity  & Confirmed	\\
145203704587705088	&	68.563651	&	+22.841887	&	CWNU\_1129	&	15 & Imaging, Other & Confirmed	\\
145220064117853696	&	69.043268	&	+22.998889	&	CWNU\_1129	&	15 & Imaging  & Confirmed	\\
3205095125321700480	&	69.400742	&	$-$02.473825	&	FSR\_1017	&	16 & Imaging  & Confirmed	\\
146764809236423808	&	69.756809	&	+23.600821	&	Theia\_7	&	187 & Imaging  & Confirmed	\\
148354733113981696	&	69.766504	&	+25.740652	&	Theia\_7	&	187 &  Imaging & Confirmed	\\
144936836795636864	&	70.002179	&	+22.350866	&	CWNU\_1129	&	15 & Imaging  & Confirmed	\\
144936836795636864	&	70.002179	&	+22.350866	&	CWNU\_1129	&	15 & Imaging & Confirmed	\\
144936836795636864	&	70.002179	&	+22.350866	&	CWNU\_1129	&	15 & Imaging  & Candidate	\\
146874275068113664	&	70.002798	&	+23.972525	&	Theia\_7	&	187	& Imaging & Confirmed\\
148401565437820928	&	70.032130	&	+26.089890	&	Theia\_7	&	187	& Imaging & Confirmed\\
3309006602007842048	&	71.626607	&	+15.472044	&	Melotte\_25	&	577 & \makecell{Radial Velocity, \\Astrometry}  & Confirmed	\\
156917493449670656	&	73.941045	&	+30.551089	&	Theia\_54	&	4 &  \makecell{Imaging, Other, \\Kinematics} & Confirmed	\\
2987545475475708416	&	74.472782	&	$-$13.703709	&	CWNU\_1018	&	153 &  Transit & Confirmed	\\
3392546632197477248	&	75.556048	&	+14.710229	&	Melotte\_25	&	577 & Imaging & Confirmed	\\
4827527233363019776	&	79.005088	&	$-$31.412694	&	HSC\_1900	&	27 &  Transit & Confirmed	\\
2955015805492793088	&	79.692355	&	$-$27.946032	&	HSC\_1900	&	27 &  Transit & Confirmed	\\
3009908378049913216	&	81.769922	&	$-$11.901175	&	FSR\_1017	&	16 &  Imaging & Confirmed	\\
3013969252448206336	&	82.469860	&	$-$09.158548	&	Theia\_6046	&	3807 & Transit  & Confirmed	\\
183489867698309888	&	83.006941	&	+36.588677	&	HSC\_1350	&	52 & Transit  & Confirmed	\\
3447802181031092992	&	83.730880	&	+31.076493	&	UPK\_369	&	135 & Transit  & Confirmed	\\
2963449441932287104	&	84.193559	&	$-$24.031219	&	HSC\_1746	&	179 & Transit  & Confirmed	\\
3404716306067442048	&	84.410417	&	+24.481036	&	Theia\_65	&	9 &  Transit & Confirmed	\\
3216110234671579648	&	84.609319	&	$-$02.678166	&	Sigma\_Orionis	&	4 &  Transit & Confirmed	\\
3216109861009561728	&	84.684311	&	$-$02.672159	&	Sigma\_Orionis	&	4 & Transit  & Confirmed	\\
3216109856714174464	&	84.685376	&	$-$02.677148	&	Sigma\_Orionis	&	4 & Imaging & Confirmed	\\
3345187475230297856	&	91.399969	&	+14.356915	&	HSC\_1553	&	39 &  Transit & Confirmed	\\
3373469800514670976	&	93.571792	&	+18.627016	&	IRAS\_06117+1901	&	18 & Transit  & Candidate	\\
3369841304767966976	&	94.535250	&	+16.023847	&	HSC\_1553	&	39 & Transit  & Confirmed	\\
3425794901747699968	&	94.535356	&	+24.833975	&	HSC\_1484	&	39 &  Transit & Candidate	\\
3007460040533835008	&	95.226318	&	$-$07.298940	&	NGC\_2215	&	450 & Transit  & Confirmed	\\
3375718851550274304	&	96.161062	&	+21.125738	&	CWNU\_1355	&	4 & Transit  & Confirmed	\\
5557593814516968960	&	104.749860	&	$-$47.023366	&	HSC\_1964	&	88 & Transit  & Confirmed	\\
2932049451436326784	&	105.149163	&	$-$20.564749	&	Tombaugh\_1	&	1252 & Transit  & Confirmed	\\
5510676828723793920	&	110.369675	&	$-$45.567737	&	Alessi\_3	&	626 & Transit  & Confirmed	\\
3047792085542514560	&	111.293211	&	$-$08.620052	&	Theia\_2299	&	222 & Transit  & Confirmed	\\
3030262468592291072	&	114.288469	&	$-$13.906670	&	NGC\_2423	&	1182 & Radial Velocity  & Confirmed	\\

\hline                                  %inserts single line
\end{tabular}
\end{table*}

\addtocounter{table}{-1}

\begin{table*}[t]
\caption{continued.}                 % title of Table
\label{table:list_stars}    % is used to refer this table in the text
\centering                        % used for centering table
\begin{tabular}{c c c c c c c}      % centered columns (3 columns)

\hline               % inserts double horizontal lines
Gaia ID Star & Right Ascension & Declination & Cluster & Age [Myr] & Discovery Method & Status  \\         % table heading
\hline                      % inserts single horizontal line
5295126507633392128	&	115.639048	&	$-$58.623272	&	UPK\_540	&	20 & Transit  & Confirmed	\\
5593780631976865280	&	117.401918	&	$-$34.451103	&	CWNU\_129	&	34 & Transit & Confirmed	\\
5290968085934209152	&	117.894946	&	$-$60.412390	&	NGC\_2516	&	125 &  Transit & Confirmed	\\
5290721997195236480	&	119.533518	&	$-$60.780499	&	NGC\_2516	&	125 &  Transit & Confirmed	\\
5290752787819629056	&	120.199911	&	$-$60.865633	&	NGC\_2516	&	125 &  Transit & Confirmed	\\
5595931482889475584	&	122.010942	&	$-$31.189361	&	ESO\_430$-$18	&	85 & Transit  & Confirmed	\\
3064530810048196352	&	123.097668	&	$-$05.768659	&	NGC\_2548	&	379 & Transit  & Confirmed	\\
5527212074861927936	&	125.901301	&	$-$42.081323	&	Theia\_1179	&	256 & Transit  & Confirmed	\\
664658700297947008	&	128.135009	&	+20.844675	&	NGC\_2632	&	346 & Transit  & Candidate 	\\
659086512807199232	&	128.245954	&	+17.306458	&	NGC\_2632	&	346 &  Transit & Candidate	\\
658832250743369344	&	128.926366	&	+16.789255	&	NGC\_2632	&	346 &  Transit & Candidate	\\
659744295638254336	&	129.362572	&	+18.976620	&	NGC\_2632	&	346 & Transit  & Confirmed	\\
664337230586013312	&	129.601086	&	+20.106007	&	NGC\_2632	&	346 & Transit  & Confirmed	\\
664292459846946560	&	129.636659	&	+19.773720	&	NGC\_2632	&	346 & Transit  & Confirmed	\\
659689109602230272	&	129.877798	&	+18.948101	&	NGC\_2632	&	346 & Transit  & Candidate 	\\
661189260077235072	&	130.009121	&	+18.949105	&	NGC\_2632	&	346 &  Transit & Candidate	\\
659494049367276544	&	130.055426	&	+18.723993	&	NGC\_2632	&	346 & Radial Velocity  & Confirmed	\\
661310756111835392	&	130.055895	&	+19.778752	&	NGC\_2632	&	346 & Transit  & Confirmed	\\
5525188767305211904	&	130.261455	&	$-$41.442796	&	Trumpler\_10	&	37 & Transit &Confirmed	\\
661302887731820416	&	130.292786	&	+19.818556	&	NGC\_2632	&	346 & Transit  & Candidate	\\
661004400386393984	&	130.343910	&	+18.933821	&	NGC\_2632	&	346 & Transit  & Confirmed	\\
660981349298550272	&	130.398549	&	+18.742998	&	NGC\_2632	&	346 &  Transit & Candidate	\\
658607950370340608	&	130.410209	&	+17.639974	&	NGC\_2632	&	346 &  Transit & Confirmed	\\
661424074528836480	&	130.432401	&	+20.226816	&	NGC\_2632	&	346 &  Radial Velocity & Confirmed	\\
661222279785743616	&	130.547744	&	+19.276957	&	NGC\_2632	&	346 & Radial Velocity  & Confirmed	\\
661222279785743616	&	130.547744	&	+19.276957	&	NGC\_2632	&	346 & Radial Velocity  & Confirmed	\\
661229701489213824	&	130.664165	&	+19.414360	&	NGC\_2632	&	346 &  Transit & Confirmed	\\
661387756285499136	&	131.207425	&	+20.177614	&	NGC\_2632	&	346 & Transit  & Candidate	\\
661167785238757376	&	131.358352	&	+19.698401	&	NGC\_2632	&	346 & Transit  & Confirmed	\\
661167785238757376	&	131.358352	&	+19.698401	&	NGC\_2632	&	346 & Transit  & Confirmed	\\
598955115237068032	&	132.486696	&	+11.692485	&	NGC\_2682	&	1688 & Radial Velocity  & Confirmed	\\
604922096120852864	&	132.753131	&	+11.886513	&	NGC\_2682	&	1688 &  Radial Velocity & Confirmed	\\
604914949295282816	&	132.753308	&	+11.814654	&	NGC\_2682	&	1688 &  Radial Velocity & 	Confirmed\\
604911375882674560	&	132.822809	&	+11.756296	&	NGC\_2682	&	1688 & Radial Velocity  & Confirmed	\\
604909657895767680	&	132.829308	&	+11.671027	&	NGC\_2682	&	1688 & Radial Velocity  & Confirmed	\\
604997240868589824	&	132.871902	&	+12.139497	&	NGC\_2682	&	1688 & Transit  & Candidate 	\\
604903468852379904	&	133.014833	&	+11.689975	&	NGC\_2682	&	1688 & Radial Velocity & Confirmed	\\
5304593027881569024	&	133.266667	&	$-$56.649444	&	OCSN\_88	&	212	& Imaging & Confirmed \\
5325454783543157760	&	134.441752	&	$-$49.000578	&	HSC\_2196	&	195 &  Transit & Confirmed	\\
5310970160975211008	&	136.527702	&	$-$54.903922	&	CWNU\_1020	&	34 & Transit  & Confirmed	\\
5256717749007641344	&	150.368272	&	$-$59.851711	&	NGC\_3114	&	166 & Transit  & Confirmed	\\
5252117594185420928	&	155.590620	&	$-$63.654620	&	HSC\_2376	&	86 &  Transit & Confirmed	\\
5251470948229949568	&	157.037281	&	$-$64.505211	&	IC\_2602	&	26 &  Transit & Confirmed	\\
5239758155778687360	&	159.157968	&	$-$64.798231	&	IC\_2602	&	26 & Transit  & Confirmed	\\
5201185574182015744	&	165.330275	&	$-$77.543975	&	Chamaleon\_I	&	6 & Imaging & Confirmed	\\
5201360671411974912	&	166.037504	&	$-$76.455369	&	Chamaleon\_I	&	6 & Imaging & Confirmed	\\
5201175987817179136	&	166.619399	&	$-$77.625873	&	Chamaleon\_I	&	6 & Imaging  &  Confirmed	\\
5201126990830255872	&	166.658187	&	$-$77.719198	&	Chamaleon\_I	&	6 & Imaging & Confirmed	\\
5201129563515179520	&	166.819463	&	$-$77.598115	&	Chamaleon\_I	&	6 & Imaging & Confirmed	\\
\hline                                  %inserts single line
\end{tabular}
\end{table*}

\addtocounter{table}{-1}

\begin{table*}[t]
\caption{continued.}                 % title of Table
\label{table:list_stars}    % is used to refer this table in the text
\centering                        % used for centering table
\begin{tabular}{c c c c c c c}      % centered columns (3 columns)

\hline               % inserts double horizontal lines
Gaia ID Star & Right Ascension & Declination & Cluster & Age [Myr] & Discovery Method & Status  \\         % table heading
\hline                      % inserts single horizontal line
5201129464729326336	&	166.907255	&	$-$77.591887	&	Chamaleon\_I	&	6 & Imaging & Confirmed	\\
5201128743176923520	&	166.942044	&	$-$77.669144	&	Chamaleon\_I	&	6 & Imaging & Confirmed	\\
5201128124701636864	&	167.013342	&	$-$77.654853	&	Chamaleon\_I	&	6 & Disk Kinematics  & 	Confirmed\\
5201128051686637568	&	167.100243	&	$-$77.658349	&	Chamaleon\_I	&	6 & Imaging & Confirmed	\\
5201128433937164416	&	167.121908	&	$-$77.655494	&	Chamaleon\_I	&	6 & Imaging & Confirmed	\\
5339389268061191040	&	167.582816	&	$-$58.982985	&	NGC\_3532	&	238 &  Transit & Confirmed	\\
5199985216726016896	&	178.769000	&	$-$79.320000	&	HSC\_2515	&	116	& Imaging & Confirmed \\
5788551695132821376	&	180.158083	&	$-$78.752300	&	HSC\_2515	&	116	& Imaging & Confirmed \\
5842480953772012928	&	186.767364	&	$-$72.451850	&	HSC\_2523	&	7 & Radial Velocity & Confirmed	\\
6117085769513415168	&	211.950000	&	$-$39.761944	&	HSC\_2636	&	10 & Transit & Confirmed	\\
1667489892685370624	&	212.720875	&	+62.522196	&	HSC\_759	&	149 &  Transit & Confirmed	\\
5790098845432414464	&	212.839201	&	$-$75.833202	&	UPK\_585	&	95 & Transit  & Confirmed	\\
5895008644342480640	&	220.003903	&	$-$52.910049	&	HSC\_2675	&	887 & Transit  & Confirmed	\\
5824848841741473024	&	230.389105	&	$-$66.264426	&	Theia\_181	&	113 &  Transit & Confirmed	\\
5201151180086109696	&	237.147011	&	$-$23.148954	&	CWNU\_1143	&	6 & Imaging & Confirmed	\\
1404488390652463872	&	237.924107	&	+52.306308	&	HSC\_759	&	149 &  Transit & Confirmed	\\
6236326362439449728	&	240.456479	&	$-$23.852297	&	OCSN\_96	&	5 & Imaging & Confirmed	\\
6243676322792317952	&	240.517939	&	$-$20.845161	&	OCSN\_96	&	5 & Imaging, Other & Confirmed	\\
6236273895118889472	&	240.713444	&	$-$24.032629	&	OCSN\_96	&	5 & Imaging & Confirmed	\\
6243332381812183296	&	240.756670	&	$-$22.131345	&	OCSN\_96	&	5 & Transit & Candidate	\\
6247526675071572224	&	240.913469	&	$-$18.858168	&	OCSN\_98	&	4 & Transit & Confirmed	\\
5997083146330022272	&	242.066835	&	$-$39.051200	&	OC\_0666	&	6 & Imaging & Confirmed	\\
5997035351934438784	&	242.226144	&	$-$39.101611	&	OC\_0666	&	6 & Imaging, Astrometry & Confirmed	\\
5997459484227732864	&	242.248082	&	$-$38.940998	&	OC\_0666	&	6 & Imaging & Confirmed	\\
6243841249531772800	&	242.376234	&	$-$21.083140	&	OCSN\_96	&	5 & Imaging & Confirmed	\\
6245758900889486720	&	242.561364	&	$-$19.319383	&	OCSN\_100	&	4 & Transit & Confirmed	\\
6049748791208799488	&	242.579956	&	$-$25.041729	&	HSC\_2907	&	10 & Imaging & Confirmed	\\
6245781097280740864	&	242.590594	&	$-$19.068497	&	OCSN\_100	&	4 & Imaging, Astrometry & Confirmed	\\
6245761404851201408	&	242.633300	&	$-$19.219055	&	OCSN\_100	&	4 & Imaging & Confirmed	\\
6049656638390048896	&	244.329167	&	$-$24.621944	&	HSC\_2907	&	10 & Transit  & Confirmed	\\
6050491992349863424	&	245.320001	&	$-$22.693647	&	HSC\_2931	&	6 & Transit  & Candidate 	\\
6025147218543620480	&	246.057741	&	$-$32.143789	&	HSC\_2907	&	10 &  Transit & Confirmed	\\
6049137943776073216	&	246.589742	&	$-$24.433610	&	HSC\_2919	&	27 & Imaging, Other & Confirmed	\\
6048935358761628288	&	246.616799	&	$-$25.446695	&	HSC\_2919	&	27 & Imaging & Confirmed	\\
6049147667582029568	&	246.860755	&	$-$24.431776	&	HSC\_2919	&	27 & Imaging & Confirmed	\\
6049146357616459136	&	246.920194	&	$-$24.483552	&	HSC\_2919	&	27 & Imaging & Confirmed	\\
5811866422581688320	&	259.356272	&	$-$66.951037	&	HSC\_2846	&	53 & Astrometry & Confirmed	\\
5949553973093167104	&	261.208614	&	$-$49.948932	&	IC\_4651	&	1665 &  Radial Velocity & Confirmed	\\
4054224522172182912	&	265.109377	&	$-$32.124087	&	NGC\_6405	&	52 & Transit  & Confirmed	\\
6702775135228913280	&	270.762500	&	$-$51.648889	&	HSC\_2846	&	53 & Astrometry  & 	Confirmed\\
6655168686921108864	&	283.274582	&	$-$50.180895	&	HSC\_2846	&	53 & Transit  & Confirmed	\\
2094015466995434624	&	283.526697	&	+37.720014	&	Stephenson\_1	&	27 & Transit  & Confirmed	\\
4184182737768311296	&	289.091824	&	$-$15.771218	&	Ruprecht\_147	&	889 & Transit & Confirmed	\\
4087782677845919744	&	289.391667	&	$-$16.871667	&	Ruprecht\_147	&	889 & Transit & Confirmed	\\
2051293083704796032	&	290.230320	&	+37.777481	&	NGC\_6791	&	3801 &  Transit & Candidate 	\\
2051293118064542976	&	290.241030	&	+37.785221	&	NGC\_6791	&	3801 & Transit  & Candidate     \\
2052827207364859264	&	290.440629	&	+38.523572	&	HSC\_572	&	103 & Transit & Confirmed	\\
6643602576214115584	&	290.712500	&	$-$54.423889	&	HSC\_2846	&	53 & Imaging & Confirmed	\\
2128198836827103616	&	293.981538	&	+46.687728	&	NGC\_6811	&	1078 & Transit & Confirmed	\\

\hline                                  %inserts single line
\end{tabular}
\end{table*}

\addtocounter{table}{-1}

\begin{table*}[t]
\caption{continued.}                 % title of Table
\label{table:list_stars}    % is used to refer this table in the text
\centering                        % used for centering table
\begin{tabular}{c c c c c c c}      % centered columns (3 columns)

\hline               % inserts double horizontal lines
Gaia ID Star & Right Ascension & Declination & Cluster & Age [Myr] & Discovery Method & Status  \\         % table heading
\hline                      % inserts single horizontal line
6752579812206787968	&	293.983181	&	$-$28.776227	&	HSC\_2846	&	53 & Imaging & Confirmed	\\
2128112181565948800	&	294.153351	&	+46.166398	&	NGC\_6811	&	1078 & Transit & Confirmed	\\
2128119878147696256	&	294.153660	&	+46.309681	&	NGC\_6811	&	1078 & Transit  & Candidate 	\\
2128146610024320768	&	294.365140	&	+46.501759	&	NGC\_6811	&	1078 & Transit  & Candidate	\\
2076485162844678784	&	295.113710	&	+40.071171	&	NGC\_6819	&	1683 & Transit  & Candidate	\\
2076294672459697536	&	295.354178	&	+40.002932	&	NGC\_6819	&	1683 & Timing & Confirmed	\\
2085198277096414464	&	299.743870	&	+45.586361	&	UBC\_143	&	500 & Transit  & Candidate	\\
2072431332214824576	&	299.809982	&	+38.393653	&	UBC\_582	&	9 & Transit  & Confirmed	\\
2082085009923554048	&	300.901460	&	+44.597328	&	NGC\_6866	&	629 & Transit  & Candidate	\\
2081884933168896640	&	301.055180	&	+44.306229	&	NGC\_6866	&	629 & Transit  & Candidate	\\
2081880913079644288	&	301.278685	&	+44.250512	&	NGC\_6866	&	629 & Transit & Confirmed	\\
2059023887421958528	&	301.564685	&	+35.588396	&	Theia\_470	&	180 &  Transit & Confirmed	\\
6845967936118138752	&	303.464715	&	$-$28.100606	&	HSC\_2846	&	53 & Imaging & Confirmed	\\
2067161441672227968	&	306.730853	&	+38.932270	&	Roslund\_6	&	168 & Transit  & Confirmed	\\
1872404225500838784	&	313.536475	&	+37.789356	&	Roslund\_7	&	175 &  Transit & Confirmed	\\
2174318981622547712	&	325.972143	&	+54.574269	&	Theia\_101	&	135 & Transit  & Confirmed	\\
6818415476100569088	&	326.272747	&	$-$20.032514	&	CWNU\_1012	&	135 & Transit  & Confirmed	\\
2003378188041736320	&	338.805332	&	+54.773557	&	Theia\_117	&	89 & Transit  & Confirmed	\\
1983367454361525888	&	338.883370	&	+45.834178	&	Alessi\_37	&	125 & Transit  & Confirmed	\\
2225320760490528384	&	340.352259	&	+69.074447	&	UPK\_230	&	123 & Transit  & Confirmed	\\
1984154395449640832	&	342.410629	&	+46.349983	&	Alessi\_37	&	125 & Transit  & Confirmed	\\
2014335027560174976	&	343.233217	&	+59.850951	&	NGC\_7429	&	78 & Transit  & Confirmed	\\
2009467558303980800	&	347.967961	&	+56.862096	&	Casado\_20	&	347 & Transit  & Confirmed	\\
6487248865942846336	&	350.720860	&	$-$61.857638	&	beta\_Tuc\_Group	&	31 & Imaging & Confirmed	\\
6387058411482257536	&	354.915467	&	$-$69.196043	&	beta\_Tuc\_Group	&	31 & Transit  & Confirmed	\\
\hline                                  %inserts single line
\end{tabular}
\end{table*}

\bibliography{Wiley-ASNA}%

\end{document}